\documentclass[conference]{IEEEtran}
\IEEEoverridecommandlockouts
\usepackage{cite}
\usepackage{amsmath,amssymb,amsfonts}
\usepackage{algorithmic}
\usepackage{graphicx}
\usepackage{textcomp}
\usepackage{xcolor}
\usepackage{multirow}

\def\BibTeX{{\rm B\kern-.05em{\sc i\kern-.025em b}\kern-.08em
    T\kern-.1667em\lower.7ex\hbox{E}\kern-.125emX}}
\begin{document}

\title{DNN-based Acoustic-to-Articulatory Inversion using Ultrasound Tongue Imaging\\
\thanks{The first and second authors were funded by Universidad Industrial de Santander (2183 grant). The last author was partially funded by the National Research, Development and Innovation Office of Hungary (FK 124584 and PD 127915 grants). The Titan X GPU used for the deep learning experiments was donated by the NVIDIA Corporation.}
}

\author{\IEEEauthorblockN{Dagoberto Porras$^1$, Alexander Sep\'ulveda-Sep\'ulveda$^1$}
\IEEEauthorblockA{
\textit{$^1$Escuela de Ingenier\'ias El\'ectrica,} \\
\textit{Electr\'onica y Telecomunicaciones,} \\
\textit{Universidad Industrial de Santander}\\
Bucaramanga, Santander, Colombia \\
dagoberto.porras@correo.uis.edu.co, fasepul@uis.edu.co}
\and
\IEEEauthorblockN{Tam\'as G\'abor Csap\'o$^{2,3}$}
\IEEEauthorblockA{\textit{$^2$Department of Telecommunications and Media Informatics,} \\
\textit{Budapest University of Technology and Economics}\\
\textit{$^3$MTA-ELTE Lend\"ulet Lingual Articulation Research Group} \\
Budapest, Hungary \\
csapot@tmit.bme.hu}
}

\maketitle

\begin{abstract}

Speech sounds are produced as the coordinated movement of the speaking organs. There are several available methods to model the relation of articulatory movements and the resulting speech signal. The reverse problem is often called as acoustic-to-articulatory inversion (AAI). 
In this paper we have implemented several different Deep Neural Networks (DNNs) to estimate the articulatory information from the acoustic signal. There are several previous works related to performing this task, but most of them are using ElectroMagnetic Articulography (EMA) for tracking the articulatory movement. Compared to EMA, Ultrasound Tongue Imaging (UTI) is a technique of higher cost-benefit if we take into account equipment cost, portability, safety and visualized structures. Seeing that, our goal is to train a DNN to obtain UT images, when using speech as input. We also test two approaches to represent the articulatory information: 1) the EigenTongue space and 2) the raw ultrasound image. As an objective quality measure for the reconstructed UT images, we use MSE, Structural Similarity Index (SSIM) and Complex-Wavelet SSIM (CW-SSIM). Our experimental results show that CW-SSIM is the most useful error measure in the UTI context. We tested three different system configurations: a) simple DNN composed of 2 hidden layers with 64x64 pixels of an UTI file as target; b) the same simple DNN but with ultrasound images projected to the EigenTongue space as the target; c) and a more complex DNN composed of 5 hidden layers with UTI files projected to the EigenTongue space. In a subjective experiment the subjects found that the neural networks with two hidden layers were more suitable for this inversion task.
\end{abstract}


\section{Introduction}
Speech sounds (=acoustics) are produced as the coordinated movement of the speaking organs (=articulation). There are several available methods to model the relation of articulatory movements and the resulting speech signal (acoustic-to-articulatory inversion~\cite{richmond2002estimating,toda2008statistical,wei2016mapping} and articulatory-to-acoustic mapping~\cite{Denby2010,Hueber2011,Csapo2017c,Toth2018}).

\subsection{Acoustic-to-Articulatory Inversion}

The articulatory movements are directly linked with the acoustic speech signal in the speech production process. The acoustic-to-articulatory inversion (AAI)~\cite{richmond2002estimating} methods estimate articulatory movements from the acoustic speech signal. Recently, there has been a significant interest in AAI, because learning the correlation between articulatory information and acoustics could improve the performance of several tasks such as speech recognition~\cite{frankel2000automatic, king2007speech}, speech synthesis~\cite{ling2009integrating}, Speaker Verification~\cite{li2016speaker}, and talking heads~\cite{wang2010synthesizing}. For Acoustic-to-articulatory inversion there are two scopes : Subject/Speaker Dependent (SD-AAI) and Independent (SI-AAI)~\cite{illa2018low}. In this paper, we deal with SD-AAI.

Several methods have been proposed to tackle the SD-AAI problem including codebooks~\cite{ouni2005modeling}, Gaussian Mixture Models (GMM's)~\cite{toda2008statistical}, Hidden Markov Models (HMM's)~\cite{zhang2008acoustic}, Mixture Density Networks~\cite{richmond2006trajectory}. Furthermore, during the past few years Deep Neural Networks (DNNs) improved accuracies in
several applications such as phone recognition~\cite{mohamed2012acoustic}. Then, researchers started to use deep architectures to accomplish an effective articulatory Inversion~\cite{uria2012deep,wu2015acoustic}, even having such success at obtaining the state-of-the-art accuracy~\cite{liu2015deep} in AAI task.

Definitely, all these approaches need parallel acoustic-articulatory data for training the AAI model. Hence, these studies are based on ElectroMagnetic Articulography (EMA) data~\cite{toda2008statistical,zhang2008acoustic,richmond2006trajectory,uria2012deep,wu2015acoustic, liu2015deep}. However, Ultrasound Tongue Imaging (UTI) is a technique of higher cost-benefit if we take into account equipment cost, portability, safety and visualized structures.

\begin{figure}
\centering
\includegraphics[trim=0.3cm 8.3cm 0.3cm 1.2cm, clip=true, width=0.51\textwidth]{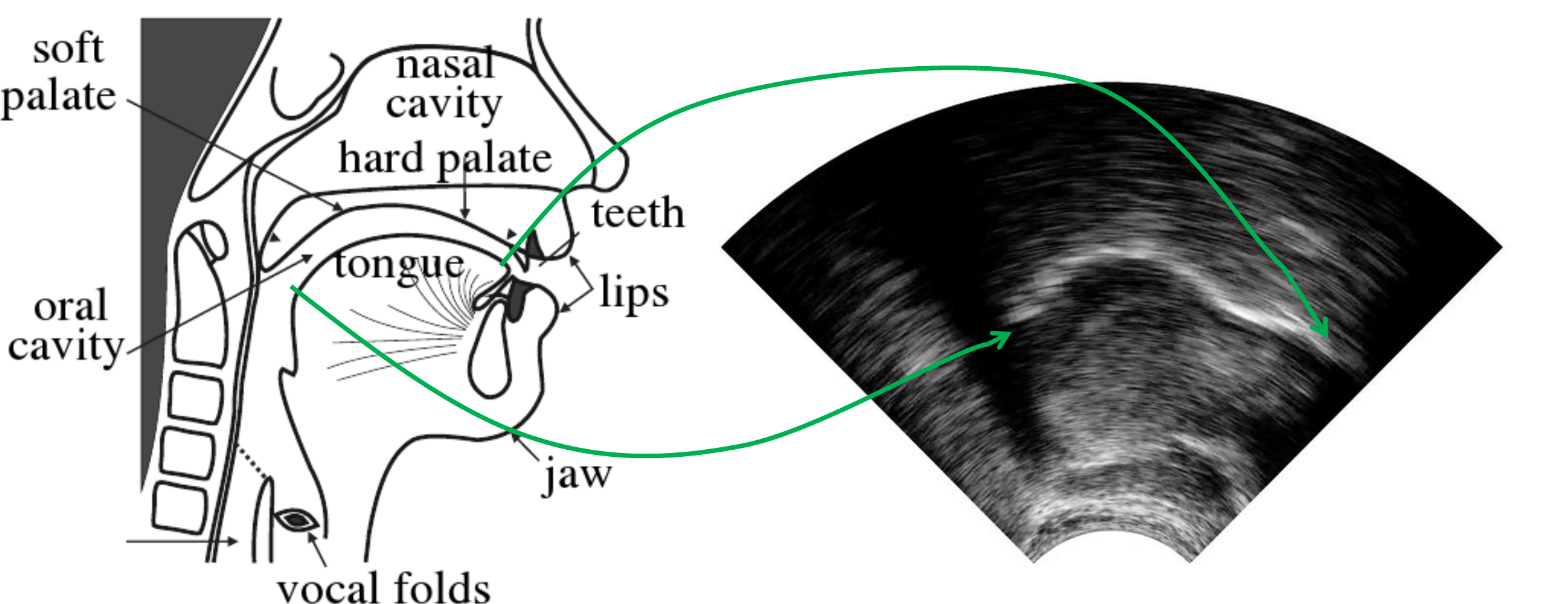}
\caption{\textit{Vocal tract (left) and a sample ultrasound tongue image (right), with the same orientation.}}
\label{fig:UTI_sample}
\end{figure}

In fact, phonetic research has employed 2D ultrasound for a number of years for investigating tongue movements during speech \cite{Stone1983, Stone2005a, Csapo2015a}. Usually, when the subject is speaking, the ultrasound transducer is placed below the chin, resulting in mid-sagittal images of the tongue movement. The typical result of 2D ultrasound recordings is a series of gray-scale images in which the tongue surface contour has a greater brightness than the surrounding tissue and air. For a guide to tongue ultrasound imaging and processing, see \cite{Stone2005a}. An example ultrasound image can be found in Fig.~\ref{fig:UTI_sample}.

\subsection{Articulatory-to-Acoustic Mapping}

In the relevant field of articulatory-to-acoustic mapping, already several types of articulatory tracking equipment types have been used. Over the last decade, there has been an increased interest in the analysis, recognition and synthesis of silent speech, which is a form of spoken communication where an acoustic signal is not produced, that is, the subject is just silently articulating without producing any sound. Systems which can perform the automatic articulatory-to-acoustic mapping are often referred to as “Silent Speech Interfaces” (SSI)~\cite{Denby2010}. Such an SSI can be applied to help the communication of the speaking impaired (e.g. patients after laryngectomy), and in situations where the speech signal itself cannot be recorded (e.g. extremely noisy environments or certain military applications). As the articulatory recording equipment, typically ultrasound tongue imaging~\cite{Denby2004,Denby2011,Hueber2010,Hueber2011,Jaumard-Hakoun2016,Csapo2017c,Grosz2018,Xu2017,Tatulli2017}, electromagnetic articulography~\cite{Wang2012a,Wang2014,Bocquelet2016,Kim2017a}, permanent magnetic articulography (PMA)~\cite{Fagan2008,Gonzalez2017a} and surface electromyography (sEMG)~\cite{Nakamura2011,Deng2014,Diener2015,Janke2017,Meltzner2017} are used. Of course, the multimodal combination of these methods is also possible~\cite{Freitas2014}, and the above methods may also be combined with a simple video recording of the lip movements~\cite{Hueber2010}.

\subsection{Deep neural networks in the inversion and mapping fields}

As deep neural networks (DNNs) became dominant in more and more areas of speech technology, such as speech recognition~\cite{Hinton2012}, speech synthesis~\cite{Ling2015} and language modeling~\cite{Arisoy2012}, it is natural that the recent studies have attempted to solve the acoustic-to-articulatory inversion and articulatory-to-acoustic conversion problems using deep learning.

In the inversion field, researchers started to deal with deep architectures to accomplish an effective articulatory inversion. Uria and his colleagues compared a deep neural network and a deep trajectory mixture density network, and obtained better inversion
accuracies with the latter than smoothing the results of a DNN~\cite{uria2012deep}. They claim to be the first demonstration of applying deep architectures to the above complex, time-varying regression problem. Wu and his colleagues used the DNN-based acoustic-to-articulatory inversion framework for a real-time speech-driven talking avatar system~\cite{wu2015acoustic}. Since the mapping between the acoustic features and the articulatory movements is non-linear, they tried four different kinds of models including
GLM, GMM, ANN and DNN to validate their ability in describing the mapping relationship; and the DNN was found to be the best. Liu et al.\ tried bidirectional LSTMs and a deep recurrent mixture density network for the AAI task~\cite{liu2015deep}, which take into account the articulatory movements as a time series, as compared to the static DNN approaches before. The advantage
of the recurrent neural network in this scenario is that it can learn proper context information on its own without the requirement of externally specifying a context window.

In the mapping field, there was a study about sEMG speech synthesis in combination with a deep neural network~\cite{Diener2015,Janke2017}. In another paper a multimodal Deep AutoEncoder was used to synthesize sung vowels based on ultrasound recordings and a video of the lips~\cite{Jaumard-Hakoun2016}. Gonzalez and his colleagues compared GMM, DNN and RNN~\cite{Gonzalez2017a} for PMA-based direct synthesis, Csap\'o et al. used DNNs to predict the spectral parameters~\cite{Csapo2017c} and F0~\cite{Grosz2018} of a vocoder using UTI as articulatory input. For the prediction of the V/U flag and F0 using articulatory input, multiple DNN architectures were compared, including DNN, RNN and LSTM neural networks~\cite{Liu2016,Zhao2017}.

According to our literature survey, in the AAI field, we only registered a single work in mapping between acoustic features of the speech signal and ultrasound images of the tongue~\cite{wei2016mapping}. They employed Deep Auto-encoders to obtain the data representations, then with a DNN framework obtained a better performance than the GMM method. They were motivated by the well known success in accuracy of deep architectures and the desire to go beyond the AAI results.  However, the study was quite limited as it only focused on Chinese vowels, i.e.\ they did not test longer speech sequences.

\subsection{Current study}

The current paper focuses on the mapping from speech acoustic features to ultrasound images. We show that simple fully connected deep neural networks are able to predict articulatory information from spectral features. Finally we evaluate our methods with objective and subjective measures.

\begin{figure*}
	\centering
	\includegraphics[width=1\linewidth]{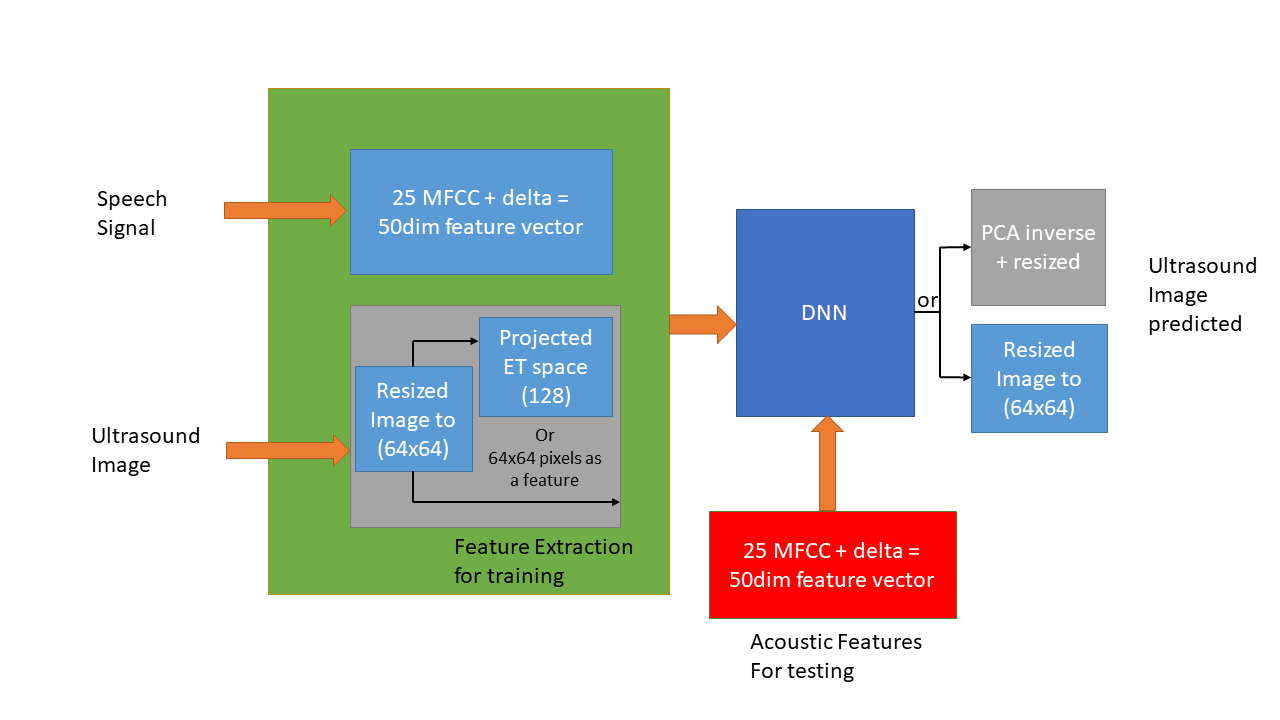}
	\caption{Framework where it is showed the two input articulatory features used for the DNN's.}
	\label{fig:DNN_framework}
\end{figure*}

\section{Methods}
\subsection{Data set}
The data acquisition setup in this work is the same as~\cite{Csapo2017c}. One Hungarian female subject (31 years old) with normal speaking abilities was recorded while reading sentences aloud from the PPSD database~\cite{olaszy2013precizios} in Hungarian language (altogether 209 sentences, of which 200 were used for training and validation; and nine for testing). The tongue movement was also recorded in midsagittal orientation using a ``Micro'' ultrasound system (Articulate Instruments Ltd.) with a 2-4 MHz / 64 element 20mm radius convex ultrasound transducer at 82 fps. During the recordings, the transducer was fixed using an ultrasound stabilization headset (Articulate Instruments Ltd.). Lip video recording was also done~\cite{Csapo2017c}, but was not used in the current study. The speech signal was recorded with an Beyerdynamic TG H56 omnidirectional condenser microphone that was mounted to the helmet approximately 5cm from the lips. Both the microphone signal and the ultrasound synchronization signals were digitized using an M-Audio – MTRACK PLUS external sound card at 44\,100~Hz sampling frequency and were resampled to 22\,050~Hz. The ultrasound and the audio signals were synchronized using the frame synchronization output of the equipment with the Articulate Assistant Advanced software (Articulate Instruments Ltd.). In the experiments below and similar to~\cite{Csapo2017c}, the raw scanline data of the ultrasound was used. 

\subsection{Ultrasound Feature Extraction}
The original raw ultrasound grayscale images of 64x842 pixels were resized to 64x64 using a bicubic interpolation. This reduction did not significantly affect the visual content of the images; instead it allows a compact size dataset, which decreases the DNNs training, synthesis time and the risk of overfitting.

\begin{figure}[ht]
	\centering
	\includegraphics[width=\linewidth]{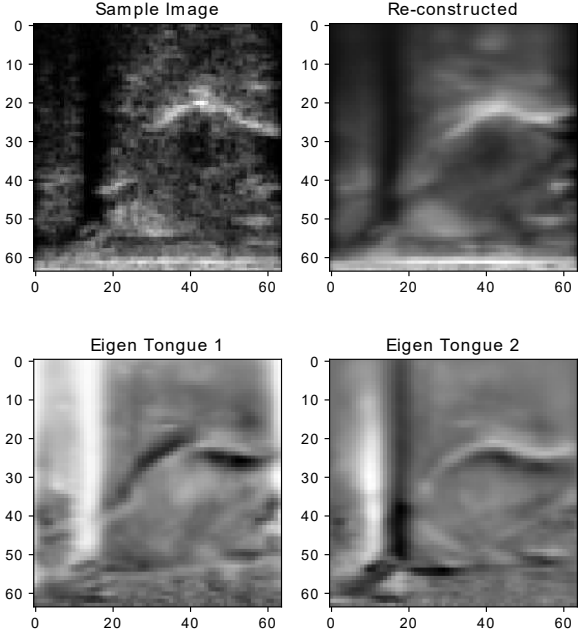}
	\caption{Reconstructed frame on the top-right from the acoustic information. Original frame is on top-left. ET$_1$ and ET$_2$ (the first two principal components out of 128) are showed in the bottom.}
	\label{fig:reconstrucedsample}
\end{figure}

In order to obtain features from the ultrasound images we tried two approaches, as can be seen in the Figure~\ref{fig:DNN_framework}. First, all the 64x64 pixels from the UTI are used as a feature. And second, we perform the 'Eigentongue' (ET) feature extraction method~\cite{hueber2007eigentongue}, which consists on finding a set of orthogonal images (\textit{EigenTongues}) that constitute a subspace for the representation of all likely tongue configurations. In practice, the eigentongues are extracted by applying PCA on the training data and the eigenvectors obtained are defined as the eigentongue. For each image, 128 ETs are extracted. Figure~\ref{fig:reconstrucedsample} shows a sample ultrasound image, one predicted by the DNN, and the first two principal ETs (out of 128), respectively. The eigentongue components extracted are supposed to encode the maximum amount of relevant information present in the images such as tongue position, hyoid bone and muscles are also present. 

\subsection{Acoustic Features}
\label{sec:MFCC}
For each 12~ms long audio frame (dictated by 82 fps ultrasound rate), 25 Mel-Frequency Cepstral Coefficients (MFCC) were extracted. Also, the first derivative $\Delta$ (known as the speed component) was included. In total an acoustic feature vector of dimension $50$ is calculated to represent the audio information~\cite{mcfee2015librosa}.

\begin{table*}
	\caption{Evaluation of the three different DNN configurations and ultrasound images features}
	\label{tab:results}
	\centering
	\begin{tabular}{ r@{	}|l	|c	*{6}{c} }
		\hline
		\multirow{2}{*}{Hidden Layers} & 
		\multirow{2}{*}{UTI features} &
		\multicolumn{2}{c}{\textbf{MSE}} & 
		\multicolumn{2}{c}{\textbf{SSIM}} &
		\multicolumn{2}{c}{\textbf{CW-SSIM}} \\
		
		& & Mean & Std.dev. & Mean & Std.dev. & Mean & Std.dev. \\	
		\hline
		$5$ x $5000$ units & $128$ ETs & $85.54$ & $11.32$ & $0.69$ & $0.09$ & $0.79$ & $0.06$ \\
		
		$2$ x $1000$ units & $64x64$ pixels & $86.02$ & $11.04$ & $0.71$ & $0.08$ & $0.81$ & $0.06$             \\
		
		$2$ x $1000$ units & $128$ ETs & $85.03$ & $11.43$ & $0.72$ & $0.09$ & $0.81$ & $0.06$             \\
		\hline
	\end{tabular}
	
\end{table*}

\subsection{DNN architecture}

The main goal of this work is to obtain ultrasound images (being the target) from the MFCC features (being the input of the neural network). For this purpose, we trained two DNN architectures: one with 5 hidden layers, with each hidden layer consisting of 5000 rectified neurons; and the other one with 2 hidden layers, each hidden layer consisting of 1000 rectified neurons. The input dimension (MFCC) is 50 (see~\ref{sec:MFCC}) and the output is different for the two approaches: 128 for the ETs and 64x64 pixels for the other hand, as Figure~\ref{fig:DNN_framework} shows. In all cases, the training criterion was the Mean Squared Error (MSE) between the predicted and the target features. In addition, we did a hyperparameter optimization to get better results; the details are in the Experimental section~\ref{sec:Experimental}. Finally, the DNN architectures are trained with 200 files containing 56\,773 ultrasound images for training and 6\,309 for validating (10\% from total number of images).

\section{Discussion}

According to our literature survey, we think that our work cannot be compared with any other research. The single study of~\cite{wei2016mapping} is the most similar work in acoustic-to-articulatory inversion with ultrasound images, however they just perform the AAI on vowel sounds. In our study we expected going beyond that approach, performing an AAI over full sentences. Although the Complex-Wavelet Structural Similarity Index (CW-SSIM) index seems to be the best quality measure for comparing ultrasound images~\cite{xu2016comparative}, we decided to include other two Quality indexes: the SSIM and the CW-SSIM indexes~\cite{sampat2009complex}.

\begin{figure*}[ht]
	\centering
	\includegraphics[width=0.5\linewidth]{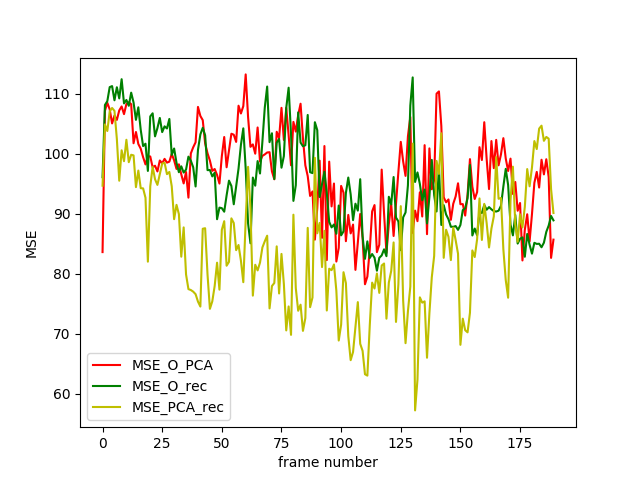}\hfill
	\includegraphics[width=0.5\linewidth]{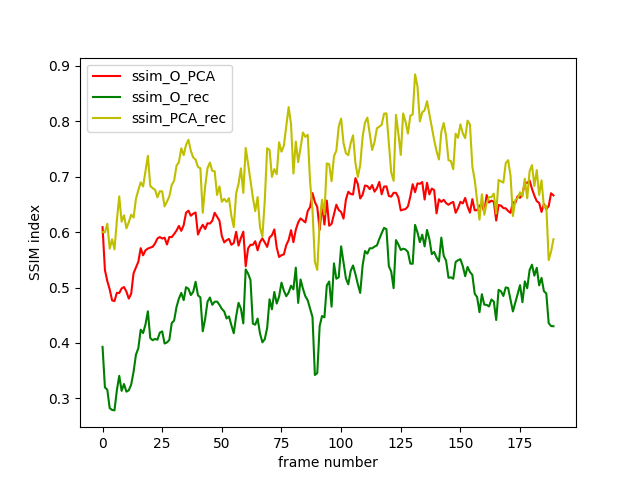}
	\includegraphics[width=0.5\linewidth]{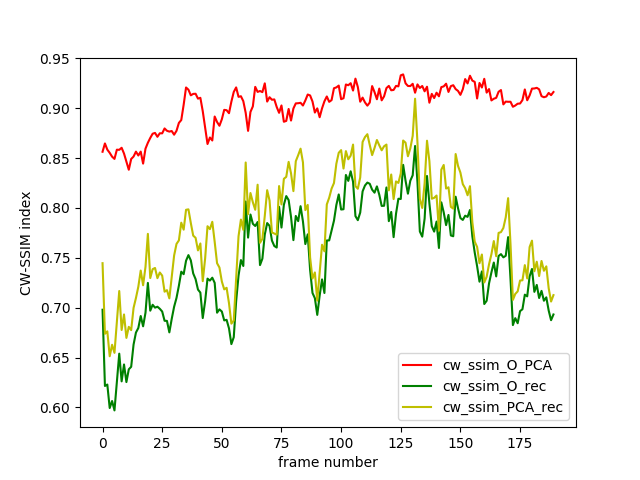}
	\caption{Image Quality measures in one test sentence. In the legend the terms before de stripe floor ("\_") means which image is the ground true, the term after the stripe floor represent the image which we are comparing. for example O\_rec represent the original image (ground truth) compared with the DNN predicted or reconstructed image. To be clear, PCA term is the original image but recovered from the PCA representation or EigenTongue space (ground truth).}
	\label{fig:Quality_measures}
\end{figure*}

\section{Experimental Results}
\label{sec:Experimental}
\subsection{Image Quality Objective Measurements}

To measure the quality of the reconstructed ultrasound images, three metrics were used. Each Quality measure is calculated over each frame, where the True image \(y\) is the resized original image and \(\hat{y}\) is the estimated image from the DNN architecture.

\begin{itemize}
	\item Mean Square Error (MSE)
\end{itemize}
\begin{equation}
MSE(y,\hat{y}) = \frac{1}{n_{samples}}\sum_{i=0}^{n_{samples}-1}(y_i -\hat{y_i})^2
\label{eq:MSE}
\end{equation}
where \(n_{samples}\) are the 64x64 pixels (4096).

\begin{itemize}
	\item The structural similarity index (SSIM)~\cite{wang2004image}, which measures three kinds of visual impact of changes in luminance \(l\), contrast \(c\) and structure \(s\) between two images.
\end{itemize}
\begin{equation}
SSIM(y,\hat{y}) = [l(y,\hat{y})]^\alpha[c(y,\hat{y})]^\beta[s(y,\hat{y})]^\gamma
\label{eq:SSIM}
\end{equation}
In our experiment the SSIM index is calculated by 11x11 circular-symmetric Gaussian weighting function, with standard deviation of 1.5 pixels. The statistics for visual impact changes are calculated as follows in~\cite{xu2016comparative}.

\begin{itemize}
	\item Complex wavelet structural similarity (CW-SSIM)~\cite{sampat2009complex} is an extension of the SSIM method to the complex wavelet domain, which is a novel image similarity measurement robust to small distortions.
\end{itemize}
\begin{equation}
CWSSIM(y,\hat{y}) = \frac{2\arrowvert\sum_{l=1}^{L}w_{y,l}w_{\hat{y},l}^{*}\arrowvert + K}{\sum_{l=1}^{L}|w_{y,l}|^2 + \sum_{l=1}^{L}|w_{\hat{y},l}|^2 + K}
\label{eq:CWSSIM}
\end{equation}

where \(w\) represents the complex wavelet coefficients of the two images. The \(^*\) indicates the complex conjugate of \(w\), and \(K\) is a small positive stabilizing constant~\cite{xu2016comparative}. In both SSIM and CW-SSIM, the value 1 means the contents in the two images compared are the same, and 0 is the minimum value (for the most diverse images).

To demonstrate the objective image quality measures, in Figure~\ref{fig:Quality_measures} we can see the three quality measures applied to one sentence from our experiment. With this figure we show that the CW-SSIM is better than the MSE and SSIM measures. For this purpose, we draw three lines for each graph. The green line is obtained by comparing the original image and the reconstructed (legend = 'O\_rec'). As a similar result, in the yellow line we compare a smooth version of the original image and the reconstructed (legend = 'PCA\_rec'). And to be coherent we expect as the best result the comparison with the original image and the smooth version in the red line (legend = 'O\_PCA').  

The red line shows what we expect to be the best result to be coherent, because we compare the original image to the recovered PCA or ET version from the original. We denote as 'smooth' version the image which is represented in the Eigen Tongue Space and then transformed to the original space (being a 64x64 pixel image). This process is done by PCA. The PCA-recovered image looks the same as the original (the same tongue contour and visualized structures) but without the speckle noise that the original image contains. So, the figure~\ref{fig:Quality_measures} suggests the CW-SSIM index could be the best measure in our experiments because it is more coherent than the other indexes. 

\subsection{Reconstruction Results}

We compared 3 systems for acoustic-to-articulatory inversion using the data of a single female Hungarian speaker. The results are shown in Table~\ref{tab:results}. Our experimental results suggest that with a simple DNN of 2 layers and 1000 units, we can get slightly better results compared to a complex 5-DNN layer of 5000 units. Moreover, it seems that is not necessary to project the ultrasound images to an EigenTongue space, because the performance is almost the same when predicting the image pixels directly. In addition, a reconstructed frame as the output of the DNN is shown in Figure~\ref{fig:reconstrucedsample} top right. Also a hyperparameter optimization was performed. Several configurations were tested including the number of neurons in the 5 hidden layers, optimizer ('rmsprop', 'adam', 'sgd') and batch size. However the results are very close to the showed in this work. Finally we think that the results presented in the Table~\ref{tab:results} are very close due to our limited amount of data. So it will be necessary to get more testing data.
\subsection{Subjective test}

\begin{figure}
\centering
\includegraphics[width=0.51\textwidth]{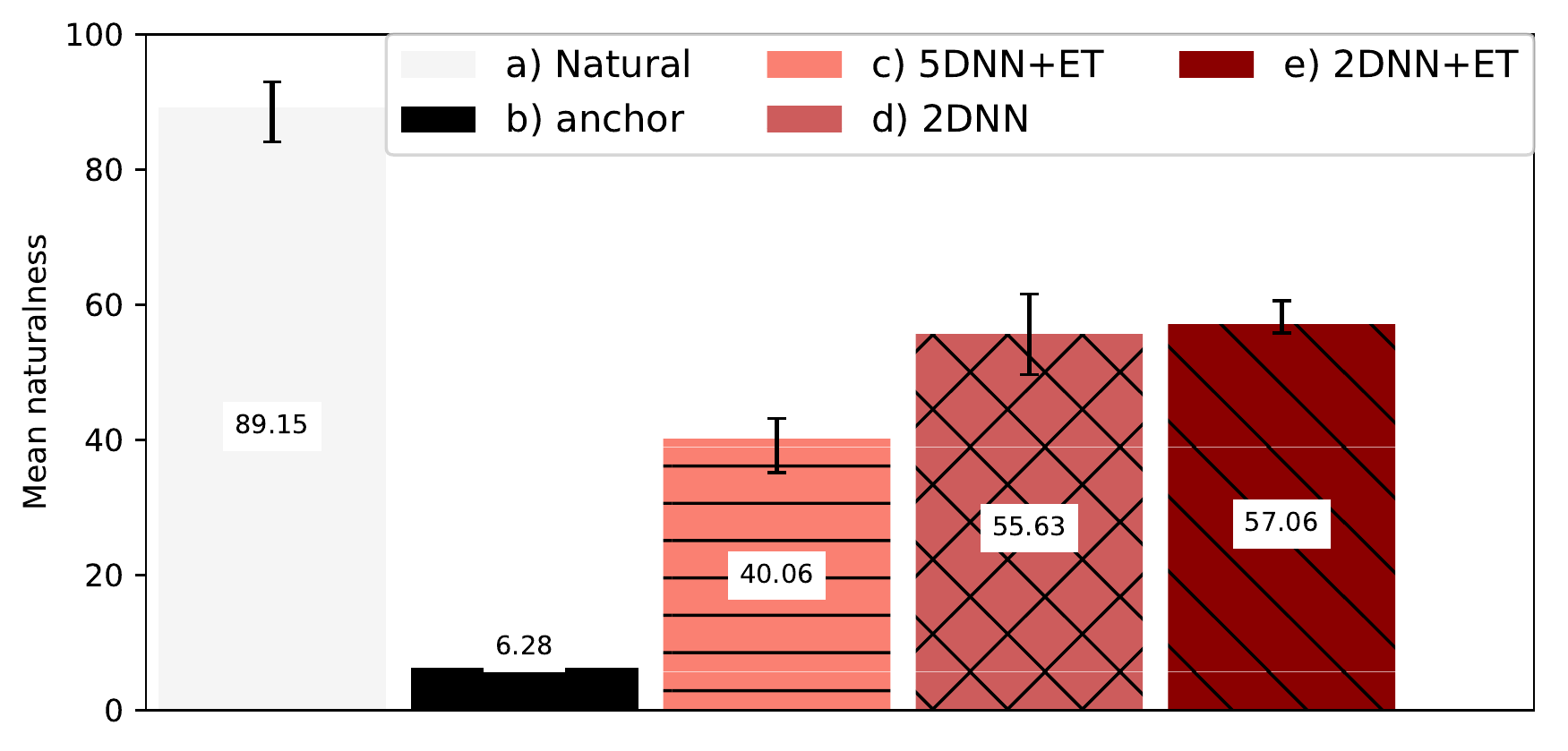}
\caption{\textit{Results of the subjective listening test concerning naturalness (synchrony of tongue ultrasound video and speech). The errorbars show the 95\% confidence intervals.}}
\label{fig:results_mushra}
\end{figure}

In order to determine which proposed system is closer to natural articulatory movement, we conducted an online MUSHRA-like (MUlti-Stimulus test with Hidden Reference and Anchor) listening test \cite{mushra}. The advantage of MUSHRA is that it allows the evaluation of multiple samples in a single trial without breaking the task into many pairwise comparisons. Our aim was to compare the natural tongue ultrasound videos with the generated videos with the various approaches and with a lower anchor. The lower anchor was a video with almost constant tongue movement. In the tests, the listeners had to rate the naturalness (i.e., whether the tongue movement is in synchrony with speech) of each stimulus in a randomized order relative to the reference (which was the PCA-compressed version of the natural sentence), from 0 (highly unnatural) to 100 (highly natural).

For the subjective test, 9 sentences were chosen which were not included in the training of the DNNs. We chose three types of the DNN configurations listed in Table~\ref{tab:results}. Together with the natural, and lower anchor samples, 45 videos were included in the test (1 speaker $\cdot$ 5 types $\cdot$ 9 sentences).

Altogether six listeners participated in the main test (two females, four males). Three of them were native speakers of Hungarian, and four of them had background in speech research. The subjects were between 23--44 years (mean: 32 years). On average the whole test took 11 minutes to complete. The MUSHRA scores of the listening test are presented in
Fig.~\ref{fig:results_mushra} for the natural, lower anchor videos and the three DNN models. In general, 'Natural' sentences should yield 100\% in MUSHRA type tests. The 'lower anchor' type ranked the lowest, as videos contained an almost constant tongue. The utterance types in which the ultrasound images were predicted based on the spectral features were ranked around 40--57\%, indicating that they are better than the lower anchor but still far away from the original tongue movement video.

The ratings of the listeners were compared by Mann-Whitney-Wilcoxon ranksum tests as well, with a 95\% confidence level, indicating that the 5DNN+ET system was significantly different from the 2DNN and 2DNN+ET systems, while the latter two were not found to be different from each other by the subjects. However, Fig.~\ref{fig:results_mushra} shows slightly better results for the DNN with 2 layers using the EigenTongues method. According to the preference of the subjects of the listening test, the '2DNN+ET' approach ranked the best, showing that a neural network with two hidden layers was more suitable for this inversion task.


When checking the videos, we also observed that the resulting image sequence is not always smooth, i.e.\ the consecutive tongue ultrasound images can contain abrupt changes. This can happen because with the above DNN, we are not taking into account the sequential nature of the articulatory and acoustic data.

\section{Conclusions}

In this work we have implemented three different kinds of AAI systems. For all cases we used deep architectures. The results over the quality measures used indicate that with a simple DNN consisting of two hidden layers can perform the best accuracy in the articulatory inversion task. However there is no much difference compared with the other system configurations, which include trying with: EigenTongue and the UTI as representation of the articulatory images. 

Moreover, we performed a hyperparameter optimization in order to obtain better results, but again we did not find a significant improvement and we think it could be due to our limited amount of data. Despite that, as a preliminary result, we can conclude that the Eigen Tongue representation procedure is not necessary.

In a similar work, Wei and his colleagues~\cite{wei2016mapping} predicted ultrasound tongue images from speech input -- using only several Chinese vowels as training data. Compared to this, our results show that the ultrasound-based acoustic-to-articulatory inversion is possible using continuous speech.

In conclusion, more testing data is necessary, and, training other deep architectures can be useful, for example Convolutional Neural Networks (CNN) \cite{McCann2017,Feigin2018}. Also we think that one could get a significant improvement including temporal information. For this purpose, we suggest using deep architectures such as RNN, LSTM or BLSTM layers. In other hand, the CW-SSIM index seems to be more coherent than the other quality measures. As a future work, we propose to compare the results presented in this work with a GMM approach and include temporal information to the features.

\section{Acknowledgements}
This work is the result of an internship done by Dagoberto Porras at Budapest University of Technology and Economics (BME). It would not have been possible with the support of Universidad Industrial de Santander (UIS), Colombia. We thank the subjects taking part in the experiments.


\bibliographystyle{IEEEtran}

\bibliography{mybib,ref_collection_csapot_nourl,mybib_reviewers}

\end{document}